# EXTRACTING MILLIMETER-WAVELENGTH RADIO EMISSION FROM A RELATIVISTIC MAGNETRON

*A.B. Batrakov, S.I. Fedotov, V.G. Korenev, O.M. Lebedenko, V.S. Mukhin, I.N. Onishchenko,*
*O.L. Rak, A.O. Shtanko, V.G. Sinitsin, M.V. Volovenko*

*National Science Center "Kharkiv Institute of Physics and Technology", Kharkiv, Ukraine*
*E-mail*: mykola.uz@gmail.com

Results of research are presented concerning operative modes of a high-voltage (relativistic) pulsed magnetron for the 8 mm wavelength range. Technical solutions are proposed for improving the output system of the device, such as to increase the efficiency of power extraction from the "field-particle" interaction space. The stability of the magnetron operation has been increased, enhanced quantitative indices of the output power to 381 kW in each of the two linear polarizations. And a stabilized frequency spectrum from 35.7 to 40 GHz of the EHF radiation generated.

## INTRODUCTION

The use of high-energy microwave radiation is necessary in a variety of applications, including plasma physics and many fields of applied research [1, 2].

The early research projects of KIPT concerned high voltage (relativistic) magnetrons (RM) intended for operation in the millimeter wave-band. The studies were carried out with the use of *Astra* type accelerators based, in their turn, on a low-impedance (about 8 Ohm), dual pulse forming line (DFL) [3]. Meanwhile, the output impedance of microwave RMs is often close to 100 Ohm. The lack of matching between the RM and the low-resistance DFL could bring forth multiple reflections along the propagation path for the radiation under study. Other unwanted effects that might (and did) occur were an increased duration of the voltages acting in the electrodynamic structure (EDS) and, consequently, a rapid deterioration of the structure [3, 4]. Altogether, many of the previous results of experimentation with different RM designs - both hornless devices and such equipped with horn antennas to allow extraction of the micro-wave radiation - now can be characterized as unsatisfactory. Indeed, the output power of those early magnetrons never was in excess of 100 kW, which figure corresponded to a lower efficiency than 0.1% [4]. Therefore, the question arose of improving the method for removal of the high frequency power from the magnetron's EDS.

## EXPERIMENTAL

The pulsed radiation generated in our early experiments was observed in the frequency range between 33 and 39 GHz, with the d.c. magnetic bias equal to $B_0$=0.5…0.9 T. Over the period of 2019 through 2021 several versions of the EDS, with 30, 40 and 48 cavities were investigated. The most complete set of results was obtained for the 48-cavity structure.

By the end of year 2023 a new accelerator was created, characterized by a noticeably higher impedance of its DFL [5]. That makes it possible to better match the pulse forming line and the RM, thus increasing the durability and operational stability of the magnetron. Currently, the research is being carried out with the use a modernized version of the installation whose schematic is shown in Fig. 1

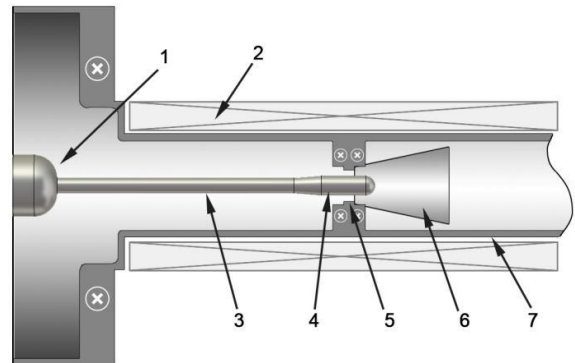

*Fig.1. Schematic of the RM:*
*1. Vacuum cavity 2. Solenoid 3. Cathode holder.*
*4. Cathode. 5. Anode. 6. Output horn.*
*7. Oversized waveguide.*

Further experiments with the RM-48 magnetron revealed microwave generation at 36 to 41 GHz with magnetic induction magnitudes $B_0$=0.35 to 0.8 T and anode voltages $U_0$=190 kV to 250 kV. The correspondent anode and cathode diameters were, respectively, $d_a$=22 mm and $d_c$=12 mm or 16 mm, while the cathode-anode gap varied like $d_{ca}$=3 mm to 5 mm. The oscillations observed could be identified as the $\pi/2$; $\pi$, or $(2/3)\pi$ modes. Yet, the net outcome of the experiments with a horn-supported axial extraction of the micro-wave power still needs to be categorized as unsatisfactory. Indeed, the RF power at the output never exceeded 100 kW, while the power level of the e-beam fed into the EDS could reach 400 MW. That meant an efficiency below 0.1 per cent, which was in a drastic contrast with the results of many numerical experiments with centimeter-wavelength RMs (like the famous A6, see, e.g., [6]) which demonstrated EHF efficiencies about 10 per cent.

In the course of experimentation aimed at increasing the level of the power extracted from the magnetron, a variety of EMS structures were manufactured and tested. Examples are shown in Figs. 2 and 3. The number and sizes of the auxiliary elements, like rods or a loop for these designs could be varied.

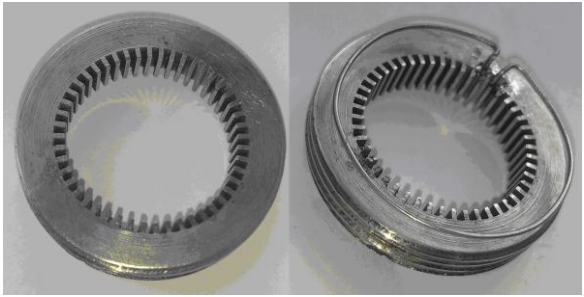

*Fig.2. Versions of power emitting structures*

This is why our further studies of EHF generation in relativistic magnetrons have become concentrated on ways to increase the amount of microwave power achievable for extraction from the anode - cathode space.

An output structure as proposed in Fig. 3 allows implementing impedance matching between the anode-cathode space, with the set of wave modes generated therein, and the output waveguide.

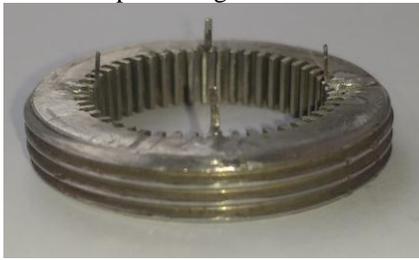

*Fig.3. Version of power emitting structure with four rod "antennas"*

Some of our recent results on microwave emission, as dependent on the d.c. magnetic field, are shown in Figs. 4 and 5 below. By applying a pulsed primary source whose impedance is matched with such of the magnetron, and implementing an electrodynamic structure of Fig. 3 (four axially oriented field extractor rods) it proved possible to increase the level of the microwave power taken off the magnetron by a factor of nearly four.

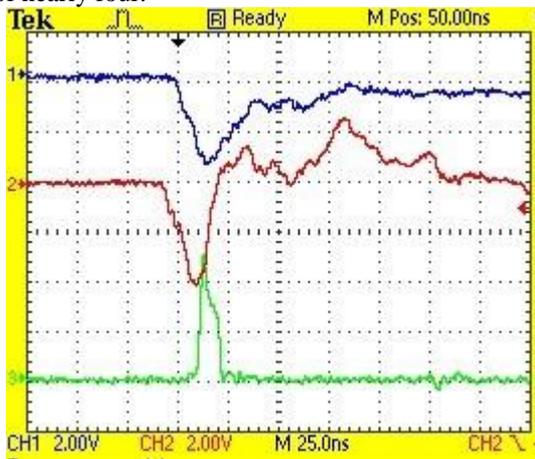

*Fig. 4. Oscillogram of microwave generation in the RM-48, experimental data:*
*1 (blue)– Anode voltage; 2 (red) – Emission current;*
*3 – (green) Signal level at the output*

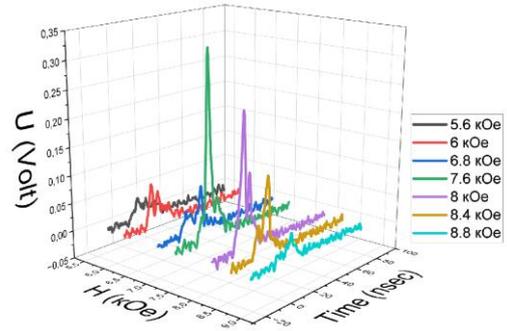

*Fig.5. Signal level versus magnetic field*

As can be seen from Fig. 5, the most efficient regimen is when the magnetic induction $B$ is close to 7.6 kOe. The oscillogram for this case is shown in Fig.4, where the blue curve represents the feed voltage, the red one is current, and the green-colored curve represents the level of microwave generation.

Since any power calculations start from estimating the frequency generated, frequency measurements were a task of top priority in our work with the RM-48. These were carried out in the order as follows:

- The operation range was established with the use of bandpass filters covering the sub-ranges like 36–40 GHz; 39–43 GHz, and 33.8–37.5 GHz.

- The accuracy of high-power measurements was estimated in a cut-off technique, using a set of waveguides of subcritical cross-section areas, plus a guide-based meter that allowed varying its cut-off frequency. As is known, the cut-off-based measuring technique exploits the ability of a single-mode guide to be operated in the capacity of a low-pass filter. The critical frequency pertinent to a waveguide is determined by the size of the guide's cross-section area. In particular, the cut-off frequency for the microwave radiation is dictated by the size of its larger wall, whence the critical wavelength can be obtained as $\lambda_{cr}=2a$.

In our experiments of 2023:

- the EHF signal was registered with a detector head built around a semi-conductor diode. The diode impedance was matched with such of the 50 Ohm load, thus allowing measurements of shorter responses from the detector. Accordingly, that enabled correct measurements of pulse-modulated signals.

- the frequencies generated in the magnetron could be measured to a noticeably better accuracy through the use of frequency mixing with the aid of a heterodyne. The block-diagram of the frequency measurements is shown in Fig. 6.

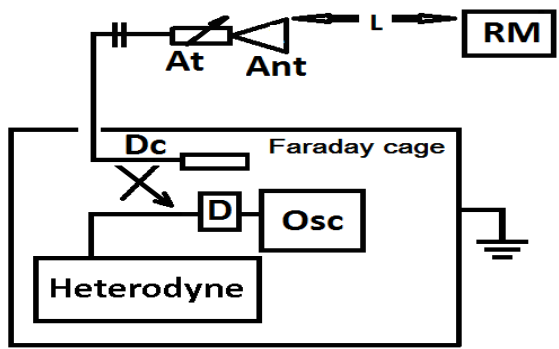

*Fig.6. Block diagram of measurements of the generated frequency:*
*RM – relativistic magnetron; Ant – horn antenna;*
*At - attenuator; D – mixing diode;*
*Osc – oscilloscope; Heterodyne;*
*Dc–Directional coupler.*

The frequency measured in a series of experiments was close to 35.7 GHz (the correspondent free-space wavelength is $\lambda$=8.4 mm).

The output power of the magnetron was measured as follows:

First, the detector head was calibrated, with the use of a power meter, at the frequency generated by the RM. The calibration data would be used as a reference in the course of further measurements.

As shown in fig. 6, the receive horn was placed in the far field zone, at a distance of $L$ = 135 cm from the exit window. A calibrated detector was used for power level measurements at the output of the receiving horn, in the maximum direction of the radiator's pattern. By adjusting the attenuators, the equality of the voltage amplitude at the detector with the previously determined reference voltage could be achieved.

The power output of the magnetron was evaluated through a set of auxiliary measurements relative such magnitudes as the attenuation ratio of the attenuator, ratio of plane areas of the measuring antenna and the output aperture of the oversized waveguide, and the shape of the directional pattern. Previous measurements have shown that the directional pattern was of a single-lobe shape. Therefore, in contrast to the case described in paper [1], the output power could be evaluated with the help of Eq. (1) below (see, e.g., [6]),

$$P_{rad} = \frac{P_r (4\pi L)^2}{G_{rad} G_r \lambda^2} \quad (1)$$

here $P_r$(W) is the power at the output of the receiving antenna (pyramidal horn);

$P_{rad}$(W) denotes the power radiated from the magnetron via an oversized circular waveguide;

$G_{rad}$ is the gain factor of the open end of the oversized waveguide;

$G_r$ is the gain factor of the pyramidal horn;

$\lambda$ (cm) stands for radiation wavelength;

$L$ (cm) is the distance from the end of the waveguide to the receiving horn.

The gain factors of the two antennas have been calculated after the equations as follows

$$G_{rad} = K_1 \left(\frac{\pi d}{\lambda}\right)^2 \quad (2)$$

$$G_r = \frac{4\pi S K_2}{\lambda^2}, \quad (3)$$

where, $K_1$ =0.84 is the aperture efficiency ratio of the cylindrical oversized waveguide;

$K_2$ =0.5 the efficiency ratio of the pyramidal horn aperture;

$S$ (cm$^2$) – the geometric area of the horn aperture.

As follows from the measurements and the results of processing after the above equations, the level of the microwave power emitted by the magnetron reaches 381 kW in each of the two linear polarizations.

## CONCLUSIONS

Measurements of the generated frequency were carried out by the heterodyne method.

In order to improve the effectiveness of energy extraction from the relativistic magnetron, the design of an EDS with four output elements of a 4 mm length was implemented. These elements were placed at the end of the cavity and allocated symmetrically.

As can be seen, the design that has been proposed offers new opportunities for enhancing the level of microwave power extraction from the magnetron.

According to the results of experimentation and calculations, the power emitted from the magnetron was 381 kW in each of the two effective linear polarizations, which is about 4 times higher than in a few preceding sets of experiments.

# ВИВЕДЕННЯ ВИПРОМІНЮВАННЯ МІЛІМЕТРОВОГО ДІАПАЗОНУ ДОВЖИН ХВИЛЬ З РЕЛЯТИВІСТСЬКОГО МАГНЕТРОНУ

*А.Б. Батраков, С.І. Федотов. В.Г. Коренєв, О.М. Лебеденко, В.С. Мухін, І.М. Оніщенко, О.Л. Рак, В.Г. Сініцин, А.О. Штанько, М.В. Воловенко*

Наведено результати досліджень робочих режимів високовольтного імпульсного магнетрону діапазону довжин хвиль 8 мм. Запропоновано шляхи подальшої модернізації системи виведення енергії з простору взаємодії "поля - частинки". Підвищено стабільність роботи магнетрону, покращено кількісні показники щодо потужності НВЧ полів на частотах від 35.7 ГГц до 40 ГГц - до рівня 381 КВт в кожній з двох поляризацій.